\newif\ifreview
\documentclass[sigconf]{acmart}

\usepackage[T1]{fontenc}
\usepackage{booktabs}
\usepackage[fleqn]{mathtools}
\usepackage{tabularx}
\usepackage{listings}
\usepackage{float}
\usepackage{fancyvrb}
\usepackage{cancel}
\usepackage[normalem]{ulem}
\usepackage{enumitem}
\usepackage{subcaption}

\AtBeginDocument{%
  }

\copyrightyear{2026}
\acmYear{2026}
\setcopyright{cc}
\setcctype{by}
\acmConference[UIST '26]{The 39th Annual ACM Symposium on User Interface Software and Technology}{November 02--05, 2026}{Detroit, MI, USA}
\acmBooktitle{The 39th Annual ACM Symposium on User Interface Software and Technology (UIST '26), November 02--05, 2026, Detroit, MI, USA}
\acmDOI{10.1145/3830398.3830624}
\acmISBN{979-8-4007-2856-3/2026/11}

\begin{document}

\title[Attribution Gradients]{Attribution Gradients: Incrementally Unfolding Citations for Critical Examination of Attributed AI Answers}

% ---- Colors ----
\definecolor{andrewpurple}{HTML}{A53DFF}
\definecolor{andreworange}{HTML}{E07400}
\definecolor{darkgreen}{HTML}{009B55}
\definecolor{darkblue}{HTML}{004d80}
\definecolor{niceblue}{HTML}{8295ff}

% ---- Annotations: no-ops for camera-ready ----
\newcommand{\andrew}[1]{}
\newcommand{\important}[1]{#1}
\newcommand{\unimportant}[1]{#1}
\newcommand{\move}[1]{#1}
\newcommand{\change}[1]{#1}
\newcommand{\new}[1]{#1}
\newcommand{\cut}[1]{}
\newcommand{\nswtm}[1]{#1}
\newcommand{\rwy}[1]{#1}
\newcommand{\keep}[1]{#1}
\newcommand{\reword}[2]{#2}
\newcommand{\hita}[1]{#1}
\newcommand{\zed}[1]{#1}
\newenvironment{changes}{}{}

% ---- Placeholders: LEFT RED ON PURPOSE, see note below ----
\newcommand{\cn}[1]{\textcolor{red}{[\emph{cite}]}}
\newcommand{\rn}[1]{\textcolor{red}{[\emph{ref}]}}
\newcommand{\fn}[1]{\textcolor{red}{[\emph{fig}]}}
\newcommand{\pn}[1]{\textcolor{red}{[\emph{(P??)}]}}
\newcommand{\fillme}[1]{\textcolor{red}{[FILL]}}
\newcommand{\factcheck}[1]{\textcolor{red}{[\emph{fact check}]}}

% ---- Fonts ----
\def\computer#1{{\small\texttt{#1}}}
\AtBeginEnvironment{quote}{\itshape}

% ---- Sectioning ----
\def\subparagraph#1{\textbf{#1.}}

% ---- URL formatting ----
\def\UrlBigBreaks{\do\/\do-\do\#}

% ---- Spacing helpers ----
\def\shortspace{\kern 0.1em}

% ---- Common strings ----
\def\KaTeX{K\kern-.2em\raisebox{.2em}{\scriptsize A}\kern-.12em\TeX}

% ---- Isotype boxes for tables ----
\def\bigbox{\color{niceblue}\rule[.25ex]{1ex}{1ex} \hskip .1ex}
\def\smallbox{\hskip .25ex \color{gray}\rule[.5ex]{.5ex}{.5ex} \hskip .25ex \hskip .1ex}
\def\boxes#1#2{
\hskip .1ex
\newcount\boxnum
\boxnum=0
\loop
\ifnum \boxnum<#1 \bigbox \else \smallbox \fi
\advance \boxnum by 1
\ifnum \boxnum<#2
\repeat
}

% ---- Inline figures ----
\newenvironment{inlinefigureenv}
{\setlength{\topsep}{2.5ex}\center}
{\endcenter}
\newcommand{\inlinefigure}[2][.5\textwidth]{%
\begin{inlinefigureenv}%
\includegraphics[width=#1]{#2}%
\vspace{-1.25ex}%
\end{inlinefigureenv}%
}

\begin{abstract}

AI answer engines are a relatively new kind of information search tool: rather than returning a ranked list of documents, they generate an answer to a search question with inline citations to sources. But reading the cited sources is costly, and citation links themselves offer little guidance about what evidence they contain. We present attribution gradients, a technique to boost the informativeness of citations by consolidating scent and information prey in place. Its first feature is bringing evidence counts, supporting/contradictory excerpts, links to sources, and contextual explanations into one place. Its second feature is the ability to unravel second-degree citations in place. In a lab study, we demonstrate usage of the full gradient in a critical reading task and its support for deep engagement that increased the depth of what readers took away from the sources versus a standard citation and document QA design.

% that with attribution gradients, participants (1) produce significantly higher-quality revisions of AI-generated claims, (2) retained evidence they encountered, and (3) used and found useful all gradient features.
\end{abstract}

\settopmatter{authorsperrow=4} 

%%
%% The "author" command and its associated commands are used to define
%% the authors and their affiliations.
%% Of note is the shared affiliation of the first two authors, and the
%% "authornote" and "authornotemark" commands
%% used to denote shared contribution to the research.
% \authornote{Both authors contributed equally to this research.}
% \authornotemark[1]
\author{Hita Kambhamettu}
\orcid{0000-0001-9620-1533}
\email{hitakam@seas.upenn.edu}
\affiliation{%
  \institution{University of Pennsylvania}
  \city{Philadelphia, PA}
  \country{USA}
}

\author{Alyssa Hwang}
\orcid{0009-0006-4827-8505}
\email{ahwang16@seas.upenn.edu}
\affiliation{%
  \institution{University of Pennsylvania}
  \city{Philadelphia, PA}
  \country{USA}
}

\author{Philippe Laban}
\orcid{0000-0001-9685-3961}
\email{plaban@microsoft.com}
\affiliation{%
  \institution{Microsoft Research}
  \city{New York, NY}
  \country{USA}
}

\author{Andrew Head}
\orcid{0000-0002-1523-3347}
\email{head@seas.upenn.edu}
\affiliation{%
  \institution{University of Pennsylvania}
  \city{Philadelphia, PA}
  \country{USA}
}

%%
%% By default, the full list of authors will be used in the page
%% headers. Often, this list is too long, and will overlap
%% other information printed in the page headers. This command allows
%% the author to define a more concise list
%% of authors' names for this purpose.
\renewcommand{\shortauthors}{Kambhamettu et al.}

%%
%% The code below is generated by the tool at http://dl.acm.org/ccs.cfm.
%% Please copy and paste the code instead of the example below.
%%
%%
%% The code below is generated by the tool at http://dl.acm.org/ccs.cfm.
%%

% These CCS concepts are standard for interactive systems. Though if you're doing something else, you probably need different concepts.

%%
%% Keywords. The author(s) should pick words that accurately describe
%% the work being presented. Separate the keywords with commas.

% \received[Revised]{29 July 2023}
% \received[revised]{12 March 2009}
% \received[accepted]{5 June 2009}

\maketitle

\begin{figure}
  \includegraphics[width=\columnwidth]{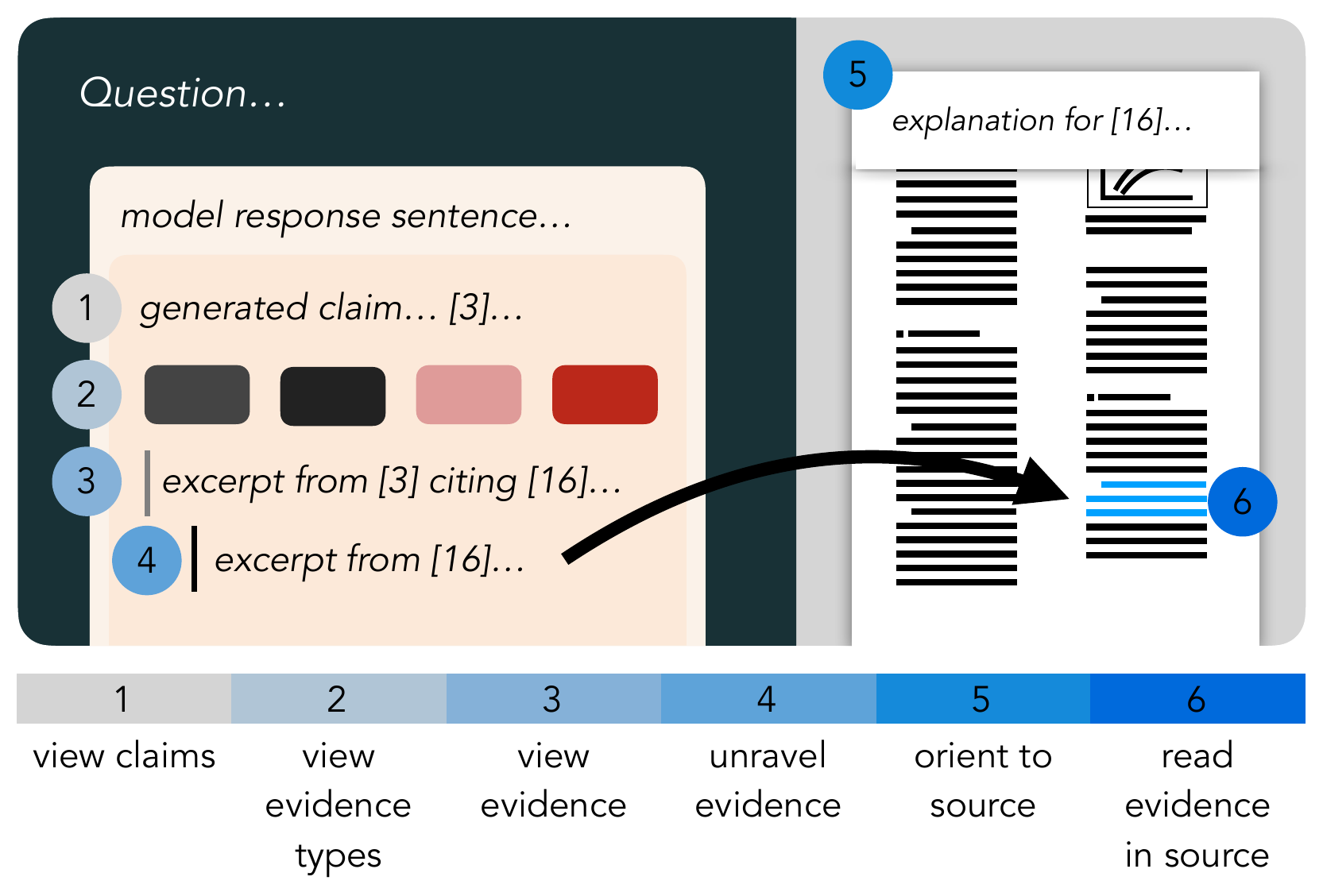}
  \vspace{-5ex}
  \caption{Attribution gradients. \textmd{When using an attribution gradient, users begin with a claim in a generated answer (1) and see color-coded clusters of evidence (2): black indicates direct support; red, direct contradiction; gray, references to sources that support the claim; and pink, references to sources that contradict the claim. Users can explore evidence excerpts (3) that might include citations to secondary references (4). Users can view excerpts in the original PDF (6) with explanations of how the evidence relates to the claim (5). This gradated unfolding of context is designed to help users access a more nuanced understanding of support for a claim.}}
  \Description{A teaser figure for attribution gradients. The figure displays an attributed AI response with numbered steps: starting with a generated claim (1), followed by color-coded evidence indicators (2), citations (3), and deep excerpts (4). An arrow connects these excerpts to a side-by-side view of the original source document (5, 6). The footer contains a six-stage progress bar: view claims, view evidence types, view evidence, unravel evidence, orient to source, and read evidence in source.
}
  % \caption{The attribution gradients interface. \textmd{Users begin with a claim in the generated answer (1) and see color-coded pieces of evidence (2): black indicates direct support; red, direct contradiction; gray, references to papers that support the claim; and pink, references to papers that contradict the claim. They can then explore evidence excerpts (3) that might include citations to secondary references (4). The interface also allows users to view these excerpts in the original PDF (6) alongside an explanation of how the evidence relates to the claim (5). This gradated unfolding of context helps users contextualize and verify each claim.}}
  % \vspace{-4ex}
  
  \label{fig:teaser}
\end{figure}

% \ifreview
% \raggedbottom
% \else
% \raggedbottom\pagebreak\flushbottom
% \fi
\section{Introduction} 

AI answer engines differ from their traditional search counterparts:
rather than returning a ranked list of documents, they generate
a written answer with inline citations attributed to source documents~\cite{venkit2024search, memon2024search}. Examples of such search engines include Perplexity~\cite{perplexity}, Microsoft Copilot~\cite{copilot}, and You.com~\cite{you}.
In these systems, the standard of presentation for citations remains a citation string:
a number or label that links to a source document.

These citations are hard to verify. Answer engines frequently
generate statements unsupported by their cited sources~\cite{venkit2024search}, surface
narrow perspectives~\cite{lindemann2024chatbots}, and reinforce biased views~\cite{sharma2024generative}. Yet users
engage with cited sources less deeply than in traditional search~\cite{venkit2024search}.
From an information foraging perspective~\cite{pirolli1995information}, the topology of a
standard cited answer is not optimal: citation strings are low-scent,
rarely conveying what evidence a source actually contains. The
cost of looking up the source compounds when a cited source itself
cites other work, what we call second-degree evidence, rather than providing direct evidence.

\change{Prior work has developed parts of the solution. For instance, the ScholarQA literature-based question answerer~\cite{Singh2025Ai2SQ} provides citation tooltips that show users source excerpts used in answer generation.} Marco~\cite{fok2024marco} lets users jump to excerpts of cited documents from tabulated answers about business document corpora. These tools reduce the cost of looking up the citations by surfacing excerpts and making it easy to jump to relevant passages.

In this work, we explore how to bring users yet more citation information
with less effort. We articulate the idea of \emph{attribution
gradients}, where context about a citation is gradually unfolded to
yield increasing information about citations under an incremental
cost structure. Our implementation of the gradient does this by letting a user to unpack a complex
sentence into atomic claims, view supporting and contradictory excerpts from sources,
jump to excerpts in their source documents, read contextual explanations
of those excerpts, and unravel citations in excerpts to reach secondary literature,
all with relatively little effort (Figure~\ref{fig:teaser}).

Most features of our gradient are known patterns from information interfaces across application areas. Evidence sorting, for instance, is a key feature of Selenite~\cite{liu2024selenite}, which lets users review counts of common themes and valences in corpora of product reviews. \change{Classification of citations into supporting and contrasting categories is a launched feature in the Scite scholarly search tool~\cite{nicholson2021scite}, which provides other classifications like the type of citing document. NEEDLE~\cite{duck2025finding} supports parallel review of a user's hypothesized claims and passages from a document corpus that support and contradict it.} Paper Plain~\cite{august2023paper} and ScholarPhi~\cite{head2021augmenting} provide refined contextual explanation widgets in scholarly papers, targeting the explanation of dense and jargon-filled scholarly passages. Scim~\cite{fok2023scim} facilitates discovery of salient document excerpts by highlighting them in place. These features have been individually validated in studies and deployments.

\begin{figure}[tb]
    \centering
    \includegraphics[width=\linewidth]{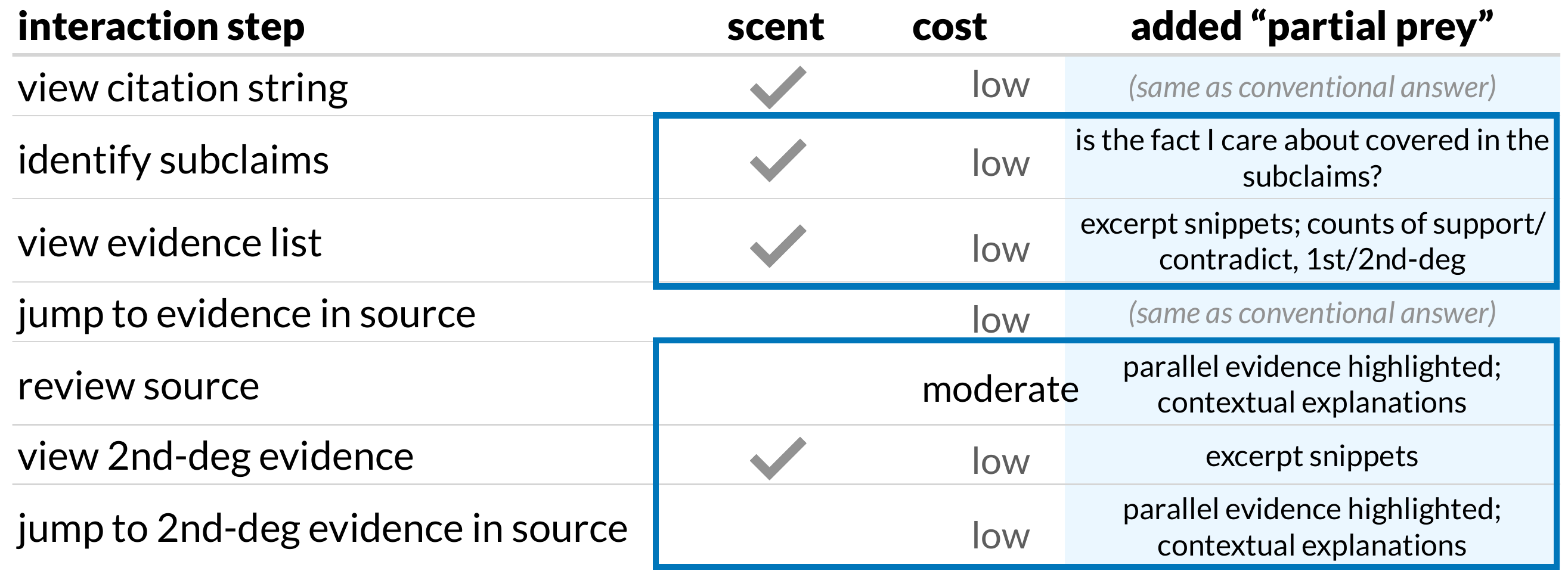}
    \caption{Articulation of support added by attribution gradients. \mdseries Compared to conventional attributed AI answers, attribution gradients change the cost structure of looking up information about evidence by introducing new scent, lesser cost, or partial prey at the steps of interaction marked by a blue box.
    % \andrew{check}
    % A comparison of attribution gradients to the conventional AI attributed answer in terms of its support for incremental scent and value, partial prey during information foraging).
    % As shown, the typical RAG QA systems do not offer much information scent at a high cost, whereas attribution gradients maintain a continuous path of low-cost, high-value interactions. The last figure also shows how attribution gradients fits into the sensemaking loop.
    % \andrew{for later, remove ``attribution gradients'' label under the table.}
    }
    \Description{The theoretical support of standard RAG and attribution gradients. Compares attribution gradients to conventional answers across interaction steps like "viewing 2nd-deg evidence". Attribution gradients are shown to provide high "scent" at "low cost" for these tasks by providing "partial prey" (snippets and highlights) that conventional answers do not support.
}
    \label{fig:ift-audit}
\end{figure}

\change{The first contribution of this work is the attribution gradients pattern. More than just a combination of past features, the idea of the gradient is to extend the citation with incremental, low-cost steps of increasing information scent and prey (Figure~\ref{fig:ift-audit}). Our second contribution is the feature of second-degree source unraveling. This part of the gradient lets users expand an excerpt's citations in place to reach evidence in the sources it cites. Our third contribution is empirical evidence of the merit of the idea. In a lab study, we explored whether attribution gradients could drive deeper engagement with attributed sources and a more nuanced understanding of generated claims. We found that (1) participants used and found useful all parts of the gradient, (2) participants produced substantially higher-quality revisions of AI-generated claims, and (3) participants retained evidence they encountered, with effort comparable to a chat-with-papers baseline. Together, this work helps sketch out designs of cited AI answers that deliver more information about sources with less effort.}

\section{Design}

\begin{figure*}
    \centering
    \includegraphics[width=\linewidth]{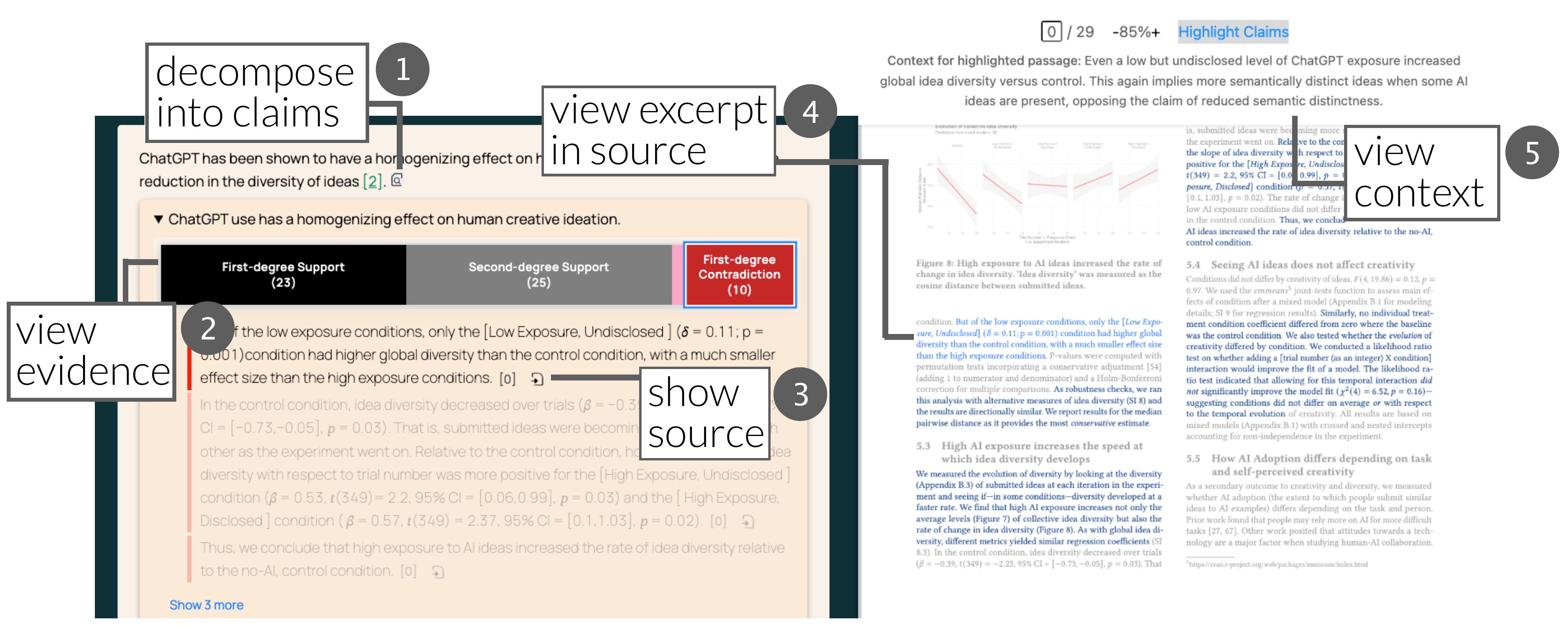}
    \caption{The full gradient. \mdseries A reader can (1) decompose a sentence in the generated answer into claims, and (2) view evidence for a particular claim. A reader can see that attributed evidence is mixed for this claim. If a reader clicks on an evidence excerpt (3), they can (4) view it in the source and (5) read context for how the evidence excerpt relates to the claim. \change{Counts shown in the evidence overview indicate the number of evidence excerpts of each type.}}
    \Description{The attribution gradients interface showing the decomposition of an AI-generated sentence into individual "inspectable claims". After clicking a magnifying glass icon next to a sentence about ChatGPT's effect on creativity, the interface reveals two atomic claims. Each claim is paired with a segmented, color-coded horizontal bar showing the distribution of evidence:
}
    \label{fig:atty-grad}
\end{figure*}

The contribution of this paper is a prototype and evaluation of \textit{attribution gradients}, which consolidate dense and rich evidence implied by a citation into an incrementally unfoldable evidence trail. The key idea is simple: rather than linking a citation directly to a source document, attribution gradients expand that single step into a series of low-cost interactions, each containing more information scent than the last. \change{We instantiate attribution gradients as an extension to OpenSciLM~\cite{asai2026synthesizing}, a scientific retrieval-augmented generation (RAG) QA system that answers questions by retrieving relevant open-access papers and generating an answer that cites them. Answers contain numerous sources, which in aggregate often tell a more complex narrative than the generated claims imply. Figure~\ref{fig:atty-grad} shows the interface, annotated for each feature below.}

\paragraph{Claim decomposition.}
A user begins by clicking on a cited sentence in the generated answer. The sentence is decomposed into atomic, testable claims, which are displayed beneath it. This gives users an entry point for looking up evidence: rather than evaluating a sentence as a whole, they can assess each claim independently.

\paragraph{Evidence overview.}
Selecting a claim displays a set of stacked bars summarizing the kinds of evidence that relate to it. \change{Our gradient draws evidence from all of the answer's citations and not just those in the current sentence, as we often found that the other citations had significant evidence to offer.} Evidence can support or contradict the claim, and can be first-degree (the source provides the evidence itself) or second-degree (the source cites other sources for the evidence). This schematizes the evidence landscape, showing not just that a claim is cited, but whether its support is strong, weak, or mixed, and direct or indirect.

\begin{figure}
    \centering
    \includegraphics[width=.9\linewidth]{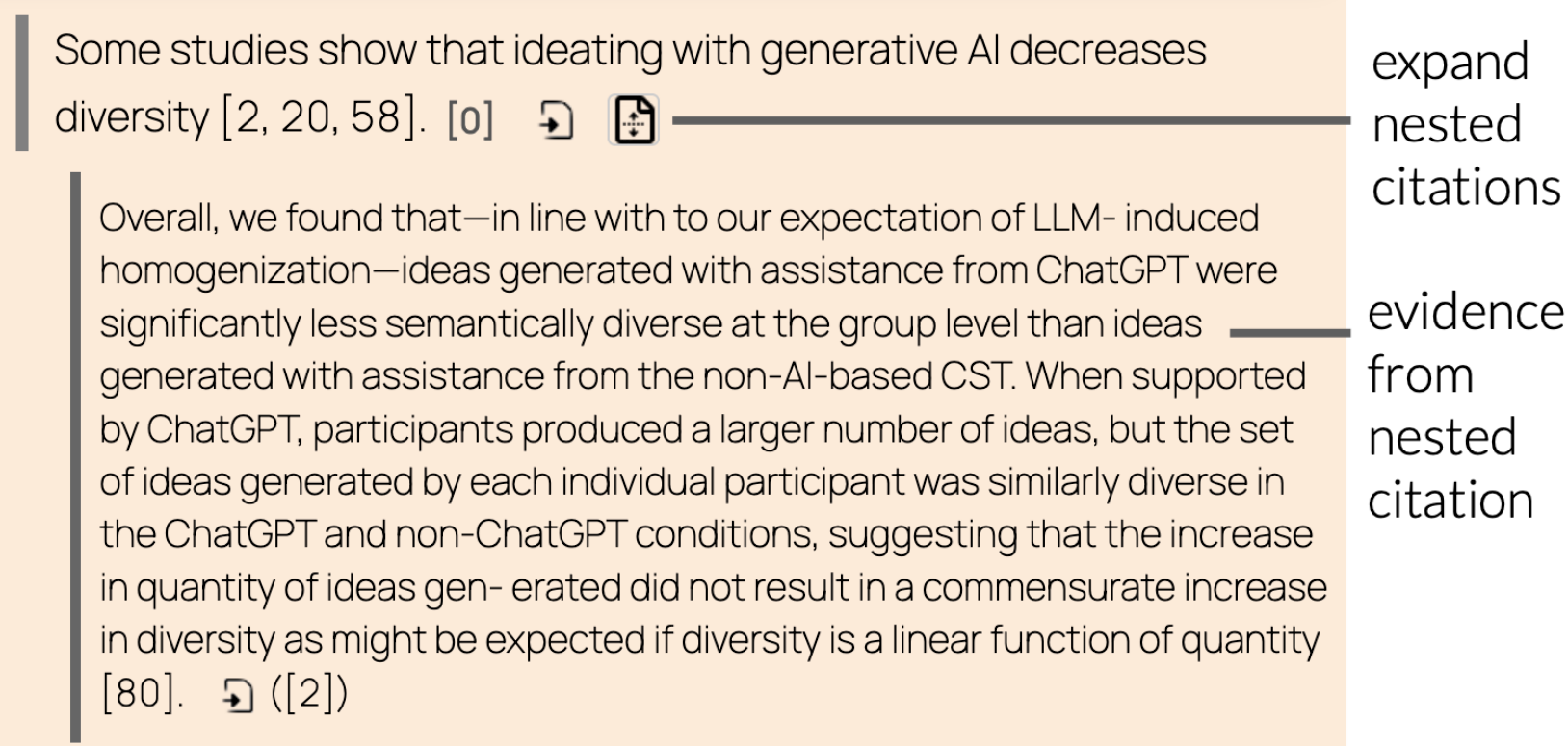}
    \caption{Unraveling citations. \mdseries If a piece of second-degree evidence cites another source for its evidence, users can sometimes unravel the citation to that source. After clicking the button to expand nested citations, excerpts from the cited source are shown that pertain to the claim. In this case, evidence from citation ``[2]'' is shown that more precisely conveys how one of the cited studies shows that ``ideating with generative AI decreases diversity.''}
    \Description{A user is unraveling citations in the attribution gradients interface. By clicking an "expand nested citations" button, the user has unraveled a citation (marked as "[2]") to view evidence from a cited source.
}
    \label{fig:second-degree}
\end{figure}

\paragraph{Evidence excerpts.}
Clicking into the overview shows the individual evidence excerpts. Excerpts are color-coded by type. This lets users scan for relevant evidence before committing to opening a source, lowering the cost of the citation link.

\paragraph{In-source viewing.}
Each excerpt has a ``show source'' button that opens the source PDF in a side panel, scrolled to and highlighting the relevant passage. Alongside it, a generated contextual explanation relates the passage to the claim. Other extracted passages from the same source are shown in a less salient color, supporting further foraging within the document.

\paragraph{Second-degree source unraveling.}
When an excerpt is second-degree, citing other sources rather than providing evidence directly, users can expand it to see excerpts from those cited sources. To our knowledge, this feature of attribution gradients has not been explored in prior sensemaking systems. It connects readers to evidence buried in the citation chain without requiring them to manually track down additional sources (Figure~\ref{fig:second-degree}).

\subsection{Implementation}
\label{sec:implementation}

To develop and test the idea of attribution gradients, we developed a prototype implementation as an extension to the OpenSciLM interface~\cite{asai2026synthesizing} (see Appendix~\ref{sec:implementation-details} for details). Given a question, the system fetches an attributed answer from \change{OpenSciLM's web API and processes it through a separate LLM pipeline (we use GPT-5.1)}: claims are extracted from cited sentences, evidence excerpts are mined from source PDFs and classified by valence (supporting or contradicting) and degree (first- or second-degree), and each excerpt is paired with a generated contextual explanation. The valence classification is similar to prior work on citation-intent classification~\cite{cohan-etal-2019-structural, lauscher-etal-2018-argument}. For second-degree excerpts, the pipeline retrieves cited references via the Semantic Scholar API~\cite{Kinney2023TheSS} and extracts excerpts from those sources. Excerpts are extracted from source PDFs using PaperMage~\cite{lo2023papermage} and highlighted in the UI via PaperCraft~\cite{papercraft}. \change{See Appendix~\ref{sec:snapshot} for a preliminary analysis of the accuracy. In our analysis, valence classification was largely accurate, and degree classification was more mixed. We note however that it is often straightforward for a user to notice misclassified degree since second-degree excerpts visibly contain citations.}

\section{Evaluation}
\label{sec:evaluation}

% \andrew{* What was the goal of the evaluation? [hint: what question do we think will be interesting to the subcommunity of people at UIST who will read this work]}

Attribution gradients are only useful if they help people interact with and learn from the sources. We sought evidence for (Q1) whether the gradients provide readers with a more nuanced understanding of the claims they read, grounded in the sources, and (Q2) whether readers used all features of the gradient, as our initial conjecture was that there was benefit to bringing them together.

We conducted a within-subjects lab study ($n=15$) comparing attribution gradients to a baseline. Both conditions shared the same OpenSciLM response view; the baseline additionally provided ChatDOC~\cite{chatdoc}, a chat-with-papers interface. This was intended as an alternative strong solution to problems attribution gradients sought to solve (QA for arbitrary questions with deep links into sources in the answers). \change{For parity with attribution gradients, ChatDOC was preloaded with all sources cited across the answer and the second-degree sources that would appear in attribution gradients}. \change{When we observed participants, their use of the baseline resembled how a user might scan and query unordered lists of sources in a passage of a Deep Research answer. Participants overwhelmingly skipped the built-in citation link in the original answer and jumped right into scanning and querying papers in ChatDOC and clicking on citations in ChatDOC's responses to read the sources.}

Participants were mostly familiar with science papers, which were the  kinds of documents they were required to consult in citations: they were doctoral (9), medical (5), and Master's (1) students recruited from a private US university, all experienced with academic literature (93\% had read over 20 papers) but intentionally varied in familiarity with the domain of the stimulus question (47\% low, 33\% moderate, 20\% high). Sessions were held over Zoom. \change{Before each task, participants received a tutorial for the interface used in that task. The experimenter demonstrated major features of the relevant interface, then asked the participant to practice those features on small practice tasks.} Interface order was counterbalanced; participants were blinded to which interface we designed.

\change{Each participant completed two 10-minute tasks, one per condition, reviewing evidence for a single claim as if preparing to explain it to a colleague. The same two study stimuli were used for all participants, with task order fixed and condition order counterbalanced.} Reviewing evidence for the claim required reconciliation across multiple sources. \change{The questions and answers that participants reviewed in the tasks were inspired by questions from ScholarQABench~\cite{asai2026synthesizing}, OpenSciLM's benchmark, lightly modified by the authors until the generated answers retrieved an interesting corpus of documents. The tasks were about recommendation algorithm effects on user preferences and ChatGPT's effect on idea diversity.} Each task pointed participants to one specific claim within a complete answer.
%; citations for other claims in the document were also processed and surfaced in attribution gradients when relevant to the focal claim.

Engagement with citations was measured through: (1) a closed-book claim revision, in which participants rewrote the claim to reflect the evidence;  (2) \change{a summative assessment of the claim-evidence relationship. This was a multiple-choice question about how the evidence relates to the claim with options ranging from ``almost certainly true'' to ``do not justify the claim'' with the ground truth answer set by an external judge;} (3) recall check questions, checking whether participants were able to recall specific pieces of evidence that they might have encountered during the task. Participants indicated whether 10 pieces of example evidence had appeared in their sources, including first-degree (8) and second-degree (2) evidence; among the options we provided three false negatives.
% An external PhD-level judge with relevant domain expertise evaluated revised claims on overall quality, evidence specificity and quantity, faithfulness, and factual correctness, blind to condition. All comparisons used the Wilcoxon signed-rank test ($\alpha=0.05$).

\subsection{Analysis}
An external PhD student with relevant domain expertise was recruited to review revised claims without knowledge of the condition in which they were written. Revised claims were evaluated on criteria of overall quality, specificity and quantity of evidence, faithfulness and correctness of the revised claim, and specificity with which the effect was described. Comparisons between conditions (for scores on revised claims, subjective Likert scores, behavioral events, and recall of evidence) were computed using the Wilcoxon signed-rank test ($\alpha=0.05$). Qualitative accounts of behavior and reactions drew from a thematic analysis of post-task questionnaires and the experimenter's structured notes.

\section{Findings}\label{results}

Below, we describe our findings. Test results are reported with test statistics and $p$-values. Qualitative data is reported with attribution to participants using pseudonyms P1--15. 

\subsection{Effect on understanding of evidence (Q1)}

Our measures suggested that participants better understood the nuance of the evidence for claims with attribution gradients and had mentally processed much of the evidence.

\subsubsection{Effect on participants' revisions to the original claim}
\label{sec:findings-revisions}

\begin{figure}
    \centering
    \includegraphics[width=\linewidth]{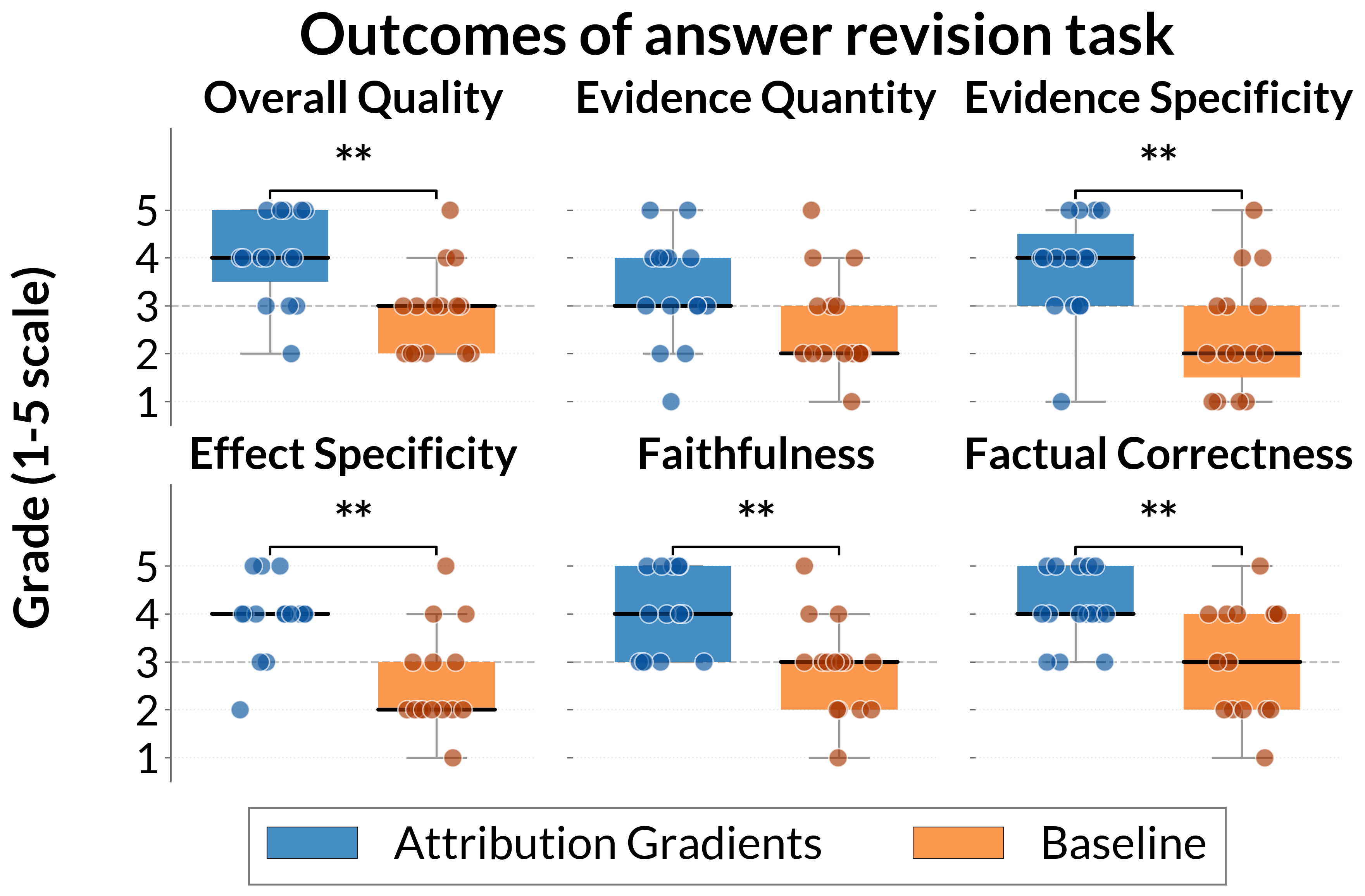}
    \caption{Comparisons of revised claims according to six quality dimensions. \mdseries \change{Participants using attribution gradients produced revisions that were rated significantly higher than the baseline on all dimensions except evidence quantity. Each box spans the interquartile range of grades, with the horizontal line marking the median. Each individual point is one participant's grade. Brackets and asterisks indicate significance from paired Wilcoxon signed-rank tests comparing conditions within each dimension (** $p < .01$).}}
    \Description{These box and whisker plots compare the quality of revised claims made by participants using attribution gradients (blue) versus a baseline condition (orange) across six dimensions: Overall Quality, Evidence Quantity, Evidence Specificity, Effect Specificity, Faithfulness, and Factual Correctness.}
    \label{fig:revision-grades}
\end{figure}

When asked to revise the claim to reflect the evidence encountered in a closed-book setting after interface use, participants produced revisions that were overall better in almost all dimensions judged. More specifically, significant differences favoring attribution gradients were found with regard to (Figure~\ref{fig:revision-grades}):

\begin{itemize}
\item Overall grade for the revised claim (M = 4.00, SD = 0.93 vs. M = 2.87, SD = 0.92; $W = 3.0$, $p = 0.004$)
\item The specificity of evidence included (M = 3.80, SD = 1.08 vs. M = 2.40, SD = 1.24; $W = 2.5$, $p = 0.006$)
\item The specificity with which the effect was described in the claim (M = 3.93, SD = 0.80 vs. M = 2.60, SD = 1.06; $W = 4.0$, $p = 0.003$)
\item The faithfulness of the revision to the available evidence (M = 4.00, SD = 0.85 vs. M = 2.87, SD = 0.99; $W = 8.0$, $p = 0.008$)
\item The factual correctness of the revised claim (M = 4.13, SD~=~0.74 vs. M = 3.07, SD = 1.16; $W = 8.0$, $p = 0.007$)
\end{itemize}

\change{To correct for multiple comparisons across the six revision-quality analyses, we applied Holm-Bonferroni correction; the five significant differences remained significant (adjusted $p \le 0.024$).}
% with large rank-biserial effect sizes ($r = 0.82$--$0.92$).}

The quantity of evidence in the revised claim had a positive but not statistically significant difference with attribution gradients (M~=~3.33, SD = 1.11 vs. M = 2.60, SD = 1.06; $W = 18.0$, $p = 0.09$). 
% The comparisons across all dimensions are shown in Figure~\ref{fig:revision-grades}.

\subsubsection{Indications of retention}
\label{sec:recall-questions}

We asked participants to recall evidence they could have come across during the task, reporting ``yes,'' ``no,'' or ``I don't recall'' for a sample of evidence (reworded so as not to be verbatim with excerpts shown in the interfaces).

Out of a possible score of 100\%, participants in the attribution gradients condition answered 51.3\% correctly ($SD=21.3$), suggesting that they had encountered, and remembered, a large amount of evidence. In other words, it is unlikely they were merely skimming the evidence. 39.3\% ($SD=17.1$) of their answers were ``I don't recall'' and these were marked wrong by default (participants only received points for a correct ``yes'' or ``no''); therefore, a 51.3\% score suggests when they commited to ``yes'' or ``no,'' they largely did so correctly. Across questions, recall was 56.0\% for questions about first-degree evidence ($SD=22.9$), 46.7\% ($SD=33.5$) for questions about second-degree evidence, and 46.7\% ($SD=30.3$) for questions where the correct answer was ``no.''
% Overall, this suggests a strong signal for better-than-random performance. 

We also asked participants in the baseline condition to recall the same pieces of evidence. This instrument was somewhat biased toward attribution gradients: evidence items were chosen through use of the attribution gradients interface (roughly uniformly sampling from top to bottom of the evidence list). This was because we mainly planned to use it as a recall check for the attribution gradients condition. \change{Our recall items were biased towards primary sources, yet participants in the baseline might have spent less time with them given that secondary sources were more numerous and the baseline offered no visual distinction between primary and secondary sources.} Still, participants could and should have come across much of this evidence in the baseline; the evidence represented key results from the sources. Baseline users were largely unable to correctly recall if these pieces of evidence were present ($M=17.3\%$, $SD=13.3$; $W=0.0, p=0.001$) for first- ($M=22.7\%$, $SD=18.3$; $W=4.0, p=0.002$) and second-degree evidence ($M=12.0\%$, $SD=14.7$; $W=0.0, p=0.005$). Participants in the baseline condition overall reported significantly less certainty about whether they had previously seen evidence, reporting ``I don't recall'' 60\% of the time ($SD=30.7$; $W=22.0, p=0.030$).

\subsection{Use of the full gradient (Q2)}
\label{sec:ag-source-engagement}

% \subsubsection{Using the full gradient}

\change{Participants' behaviors and questionnaire responses helped clarify how participants used features of the gradient, both individually and in tandem, during the claim review task.} In general, participants reported the features of attribution gradients as highly useful. To calibrate, all features below were rated on average as at least 6 out of 7, while for comparison, participants rated features in the baseline below 5 (ability to ask questions of papers, $M = 4.71$, $SD = 1.59$; ability to jump to passage from answer, $M = 4.79$, $SD = 1.93$).

\paragraph{Evidence overview.} 
\label{sec:evidence-facets}
All participants filtered evidence types at least once when using attribution gradients. 15 toggled to view first-degree support, 13 to view second-degree support, and 7 to view first-degree contradictions (second-degree contradictions were not available in the provided stimuli).
Participants reported the sorting of excerpts as highly useful, both along the dimensions of first- and second-degree evidence ($M=6.79, SD=0.58$) and supporting and contradictory valence ($M=6.14, SD=1.17$). 

As P11 described, ``in [the baseline], the onus was on me to generate questions from the contradictory perspective\ldots{} [with attribution gradients] it was easier for me to see both sides of the claim.'' Overall, we saw some indications that participants may have engaged more meaningfully with different valences of information with attribution gradients: of the eight participants whose revisions incorporated both supporting and contradictory evidence,\footnote{\change{As assessed by an independent external grader during review of revised claims.}} \change{six} had done so using attribution gradients. We note that all six of these participants had clicked to filter to contradictory evidence or open a contradictory excerpt during the task.

\paragraph{Second-degree source unraveling}
\label{sec:unraveling-evidence}
When using attribution gradients, all 15 participants viewed at least one excerpt containing second-degree evidence, and 12 clicked to expand at least one second-degree excerpt to show excerpts from the sources it cited. P8 described that the second-degree excerpts helped them ``go down more rabbit holes\ldots{} if I wanted to know exactly what study was done.'' Participants reported the ability to unravel citations to view original sources to be very useful ($M = 6.29, SD = 0.99$). This feature could support access to quite a lot of secondary sources; P4, for instance, opened seven sources providing second-degree evidence.

\paragraph{In-source viewing}
The ability to open excerpts in the source was frequently exercised in the claim review task. Participants opened an excerpt in source a median of 20 times ($SD = 11.24$). Participants reported this feature as highly useful as well ($M = 6.79, SD = 0.58$). Participants' descriptions aligned with the intent of the feature: P10 told us that with this feature they ``found myself reading the source itself'' and P4 told us that using the feature ``made me more confident in citing the sources since I had more opportunities to read more papers and also more of each paper.'' The source viewer also appeared to support another mode of foraging: six participants, for instance, clearly engaged with content in a paper aside from the excerpt they originally opened the paper to view. Sometimes they scrolled to see additional content, and sometimes they clicked on different highlighted excerpts in the source.

\paragraph{Use of features together}
Many participants reviewed lists in a way that coordinated features. For instance, we often saw a pattern resembling incremental, iterative processing of  available evidence. Nine participants opened excerpts from the evidence overview in quick succession, viewing each briefly and working through on average about 50\% of the list of excerpts that were displayed. P4 described the utility of this approach: ``each amount of information [in the evidence excerpt] was small so I was able to quickly filter out which info was useful to me\ldots{} I liked being able to be given all of the information so I can make that decision [of what to read].'' 
% Meanwhile, it also appeared that incremental, iterative processing was useful to more than just the excerpts. 

% Consider the experience of P9, whose was ultimately able to revise their claim to include synthesis of both supporting and contradictory evidence. They began the task opening nine excerpts from the same paper. When they opened a tenth excerpt, now from a different paper, P9 remarked that ``this study seems irrelevant to the claim'' and toggled away from that document, instead beginning to review contradictory evidence.

\medskip

In addition to the features above, we validate use of the other features. By nature of interacting with the system at all, participants had descended into \emph{claim decompositions}. By nature of opening sources, all participants had engaged with \emph{evidence excerpts}.

\section{Discussion}
\label{sec:discussion}

Our results suggest that structuring attributions as a gradient supports deeper engagement with the available evidence. \change{Because we evaluated the gradient as a single integrated design, however, our results speak to the value of the assembled gradient rather than to any individual feature within it.} We see four aspects of this design as potentially generative for attributed AI interfaces more broadly: \textit{incrementality} (breaking review into smaller steps), \textit{schematization} (pre-loading productive evidence schemas), \textit{recursion} (treating sources as launching points into further literature), and \textit{context} (providing orientation for decontextualized excerpts). These dimensions are not new to sensemaking research, but handling them well in attributed AI answers may change how users engage with generated content that is linked back to sources. That said, our evidence comes only from scientific QA with readers experienced in academic literature; whether the pattern transfers to other settings is an open question we take up below.

\subsection{Future studies}
\label{sec:limitations}

Our study might inspire future ones to be run across:

% Version without parens and updated last sentence
\paragraph{Different AI search engines.} \change{In principle, attribution gradients could apply wherever an answer provides citations, claims are in contention, and sources are available (e.g. Google AI Overviews and Deep Research~\cite{deepresearch}). Their efficacy may vary with source availability, source count, and evidence depth. Processing citations comes at a cost, and search engine builders will have to decide how many citations to process, how thoroughly.}

% \paragraph{Different AI search engines.} \change{In principle, attribution gradients could apply wherever an answer provides citations, claims are in contention, and sources are available (e.g. Google AI Overviews and Deep Research~\cite{deepresearch}). Their efficacy may vary with source availability (limited for paywalled literature), source count (potentially overwhelming for deep research tools), and evidence depth (likely shallower in news, where second-degree chains are rare). Building our prototype also surfaced a tension between preprocessing latency and gradient quality that interactive settings could sharpen.}

\paragraph{Different participants.} \change{Are attribution gradients still useful when readers have more knowledge of claims and sources?}

% \paragraph{Different backend abilities.} \change{As AI search engines become more accurate, will there still be a need for close review of citations?}

\paragraph{Different subsets of the gradient} We evaluated the full gradient. Which of its parts most benefited users? \change{Follow-on work could isolate its individual parts, and imperfections in them (e.g., how misclassified evidence affects readers' conclusions, and whether readers detect and recover from such errors).} % We see this as important, given the positive results from our study.

\subsection{Future systems}

We see several areas for fruitful future systems work.

\paragraph{Richer schemas} Attribution gradients could be expanded beyond the binary support-versus-contradict categorization to incorporate richer, multi-dimensional contextual relationships between claims and evidence. Users might glean more meaning that goes beyond whether evidence agrees or disagrees with a claim.

\paragraph{Richer evidence representations} If raw data, such as tables or numeric results, can be extracted from sources, then attribution gradients could provide greater context about the relationship of this data to claims. A gradient could incorporate information about whether statistical tests could be replicated, the uncertainty of a result, how a statistic was computed, or issues in a logical argument.

\paragraph{Richer evidence pools} \change{Our gradients draw only on the sources cited in the original answer, so they help readers assess whether the answer represents its own citations. If the AI answer engine selects sources unevenly, the evidence shown in the overview inherits that bias. A fully critical future instantiation could do a post-hoc search for evidence on all sides of a claim before constructing the gradient.}

% \paragraph{Tracking conclusions} \change{The gradient does not help readers keep track of their takeaways from evidence excerpts. Future systems could support this, perhaps by allowing users to filter excerpts that they do not find useful or highlight excerpts that are useful.}

\section{Conclusion}

We introduce attribution gradients to support the close reading of attributed AI answers. The gradient consolidates information about the claim, the valence and nature of the evidence, evidence excerpts, evidence chains, and contextual information, into one view.
% Our contribution is an instantiation of this consolidation, the novel subfeature of second-degree source unraveling, and a first study examining its effect on evidence review.
% , attribution gradients establish tight interconnections among answer, claim, excerpt, and context.
% By instantiating this in an attributed scientific QA system,
Our study showed participants revising claims in a way that suggested having acquired more information from attribution gradients than in the baseline.
% that participants  that participants engaged with more sources and produced higher quality claim revisions compared to a baseline.
This work demonstrates how reframing citation inspection as a gradient of increasing information scent enables users to make better use of attributed AI answers.

\begin{acks}
% We thank the authors of augmented formulas that inspired examples used in this paper and in our study materials, including \citet{mohammed2020continuous,ref:murad2020navierstokes,W4,ref:cockett2016pixels}, and \citet{ref:hohman2019gamut}.
% We thank Dr. Rowan Cockett for his feedback on tool design and implementation. We thank members and friends of \emph{Penn HCI} for their frequent feedback on early prototypes and study designs, as well as creators of open source tools like KaTeX for making the construction of FFL possible.

We thank Princess Sampson, Mialy Rasetarinera, and Xiaorui Liu for grading participant responses, and Litao Yan for his feedback on figures and study design.
This research was developed with funding from the Defense Advanced Research Projects Agency's SciFy program (Agreement No. HR00112520300). The views expressed are those of the authors and do not reflect the official policy or position of the Department of Defense or the U.S. Government.

\end{acks}

\bibliographystyle{ACM-Reference-Format}
\bibliography{refs,andrew-base,andrew-extras}

@preamble{"
\def\articleno#1{Article #1}
\def\paperno#1{Paper #1}
\def\toappear{To appear}
\def\eatnexttoken#1{}
\def\pubmit{The MIT Press}
\def\pubieee{IEEE}
\def\pubacm{ACM}
\def\pubwiley{John Wiley \& Sons, Ltd.}
\def\pubspringer{Springer}
\def\pubelsevier{Elsevier}
\def\pubmorgankaufmann{Morgan Kaufmann}
\def\pubemerald{Emerald Group Publishing Limited}
\def\journalspae{Software---Practice and Experience}
\def\journalsoftware{IEEE Software}
\def\journalcacm{Communications of the ACM}
\def\journaljvlc{Journal of Visual Languages and Computing}
\def\journalmms{International Journal of Man-Machine Studies}
\def\journaltosem{ACM Transactions on Software Engineering and Methodology}
\def\journaltse{IEEE Transactions on Software Engineering}
\def\journaltochi{ACM Transactions on Computer-Human Interaction}
\def\journalsigplan{ACM SIGPLAN Notices}
\def\journaltog{ACM Transactions on Graphics}
\def\journaltvcg{IEEE Transactions of Visualization and Computer Graphics}
\def\confinteract{Proceedings of the International Conference on Human-Computer Interaction}
\def\confoopsla{Proceedings of the Conference on Object-Oriented Programming Systems, Languages, and Applications}
\def\confpldi{Proceedings of the Conference on Programming Language Design and Implementation}
\def\confcscw{Proceedings of the Conference on Computer-Supported Cooperative Work and Social Computing}
\def\confpopl{Proceedings of SIGPLAN Symposium on Principles of Programming Languages}
\def\conficsme{Proceedings of the International Conference on Software Maintenance and Evolution}
\def\conffsesymposium{Proceedings of the International Symposium on Foundations of Software Engineering}
\def\conffsemeeting{Proceedings of the Joint Meeting on Foundations of Software Engineering}
\def\conficsm{Proceedings of the International Conference on Software Maintenance}
\def\confcgit{Proceedings of the International Conference on Computer Graphics and Interactive Techniques}
\def\confaaai{Proceedings of the AAAI Conference on Artificial Intelligence}
\def\confchi{Proceedings of the CHI Conference on Human Factors in Computing Systems}
\def\confemnlp{Proceedings of the Conference on Empirical Methods in Natural Language Processing}
\def\confiui{Proceedings of the International Conference on Intelligent User Interfaces}
\def\confuist{Proceedings of the Symposium on User Interface Software and Technology}
\def\confase{Proceedings of the International Conference on Automated Software Engineering}
\def\confissta{Proceedings of the International Symposium on Software Testing and Analysis}
\def\conffase{Proceedings of the International Conference on Fundamental Approaches to Software Engineering}
\def\conficpc{Proceedings of the International Conference on Program Comprehension}
\def\conficse{Proceedings of the International Conference on Software Engineering}
\def\confchase{Proceedings of the International Workshop on Cooperative and Human Aspects of Software Engineering}
\def\confacl{Proceedings of the Annual Meeting of the Association for Computational Linguistics}
\def\confvlhcc{Proceedings of the Symposium on Visual Languages and Human-Centric Computing}
\def\conflas{Proceedings of the Conference on Learning at Scale}
\def\confisese{Proceedings of the International Symposium on Empirical Software Engineering}
\def\confsigcse{Proceedings of the Technical Symposium on Computer Science Education}
\def\confwcre{Proceedings of the Working Conference on Reverse Engineering}
\def\confhatra{Proceedings the Workshop on Human Aspects of Types and Reasoning Assistants}
\def\confonward{Proceedings the Symposium on New Ideas in Programming and Reflections on Software}
\def\confvldb{Proceedings of the International Conference on Very Large Data Bases}
"}

@preamble{"
\def\articleno#1{Article #1}
\def\paperno#1{Paper #1}
\def\toappear{To appear}
\def\eatnexttoken#1{} 
\def\pubmit{The MIT Press}
\def\pubieee{IEEE}
\def\pubacm{ACM}
\def\pubwiley{John Wiley \& Sons, Ltd.}
\def\pubspringer{Springer}
\def\pubelsevier{Elsevier}
\def\pubmorgankaufmann{Morgan Kaufmann}
\def\pubemerald{Emerald Group Publishing Limited}
\def\journalspae{Software---Practice and Experience}
\def\journalsoftware{IEEE Software}
\def\journalcacm{Communications of the ACM}
\def\journaljvlc{Journal of Visual Languages and Computing}
\def\journalmms{International Journal of Man-Machine Studies}
\def\journaltosem{ACM Transactions on Software Engineering and Methodology}
\def\journaltse{IEEE Transactions on Software Engineering}
\def\journaltochi{ACM Transactions on Computer-Human Interaction}
\def\journalsigplan{ACM SIGPLAN Notices}
\def\journaltog{ACM Transactions on Graphics}
\def\journaltvcg{IEEE Transactions of Visualization and Computer Graphics}
\def\confinteract{Proceedings of the International Conference on Human-Computer Interaction}
\def\confoopsla{Proceedings of the Conference on Object-Oriented Programming Systems, Languages, and Applications}
\def\confpldi{Proceedings of the Conference on Programming Language Design and Implementation}
\def\confcscw{Proceedings of the Conference on Computer-Supported Cooperative Work and Social Computing}
\def\confpopl{Proceedings of SIGPLAN Symposium on Principles of Programming Languages}
\def\conficsme{Proceedings of the International Conference on Software Maintenance and Evolution}
\def\conffsesymposium{Proceedings of the International Symposium on Foundations of Software Engineering}
\def\conffsemeeting{Proceedings of the Joint Meeting on Foundations of Software Engineering}
\def\conficsm{Proceedings of the International Conference on Software Maintenance}
\def\confcgit{Proceedings of the International Conference on Computer Graphics and Interactive Techniques}
\def\confaaai{Proceedings of the AAAI Conference on Artificial Intelligence}
\def\confchi{Proceedings of the CHI Conference on Human Factors in Computing Systems}
\def\confemnlp{Proceedings of the Conference on Empirical Methods in Natural Language Processing}
\def\confiui{Proceedings of the International Conference on Intelligent User Interfaces}
\def\confuist{Proceedings of the Symposium on User Interface Software and Technology}
\def\confase{Proceedings of the International Conference on Automated Software Engineering}
\def\confissta{Proceedings of the International Symposium on Software Testing and Analysis}
\def\conffase{Proceedings of the International Conference on Fundamental Approaches to Software Engineering}
\def\conficpc{Proceedings of the International Conference on Program Comprehension}
\def\conficse{Proceedings of the International Conference on Software Engineering}
\def\confchase{Proceedings of the International Workshop on Cooperative and Human Aspects of Software Engineering}
\def\confacl{Proceedings of the Annual Meeting of the Association for Computational Linguistics}
\def\confvlhcc{Proceedings of the Symposium on Visual Languages and Human-Centric Computing}
\def\conflas{Proceedings of the Conference on Learning at Scale}
\def\confisese{Proceedings of the International Symposium on Empirical Software Engineering}
\def\confsigcse{Proceedings of the Technical Symposium on Computer Science Education}
\def\confwcre{Proceedings of the Working Conference on Reverse Engineering}
\def\confhatra{Proceedings of the Workshop on Human Aspects of Types and Reasoning Assistants}
\def\confonward{Proceedings the Symposium on New Ideas in Programming and Reflections on Software}
\def\confvldb{Proceedings of the International Conference on Very Large Data Bases}
"}

@preamble{"
\def\articleno#1{Article #1}
\def\paperno#1{Paper #1}
\def\toappear{To appear}
\def\eatnexttoken#1{} 
\def\pubmit{The MIT Press}
\def\pubieee{IEEE}
\def\pubacm{ACM}
\def\pubwiley{John Wiley \& Sons, Ltd.}
\def\pubspringer{Springer}
\def\pubelsevier{Elsevier}
\def\pubmorgankaufmann{Morgan Kaufmann}
\def\pubemerald{Emerald Group Publishing Limited}
\def\journalspae{Software---Practice and Experience}
\def\journalsoftware{IEEE Software}
\def\journalcacm{Communications of the ACM}
\def\journaljvlc{Journal of Visual Languages and Computing}
\def\journalmms{International Journal of Man-Machine Studies}
\def\journaltosem{ACM Transactions on Software Engineering and Methodology}
\def\journaltse{IEEE Transactions on Software Engineering}
\def\journaltochi{ACM Transactions on Computer-Human Interaction}
\def\journalsigplan{ACM SIGPLAN Notices}
\def\journaltog{ACM Transactions on Graphics}
\def\journaltvcg{ACM Transactions of Visualization and Computer Graphics}
\def\confinteract{Proceedings of the International Conference on Human-Computer Interaction}
\def\confoopsla{Proceedings of the Conference on Object-Oriented Programming Systems, Languages, and Applications}
\def\confpldi{Proceedings of the Conference on Programming Language Design and Implementation}
\def\confcscw{Proceedings of the Conference on Computer-Supported Cooperative Work and Social Computing}
\def\confpopl{Proceedings of SIGPLAN Symposium on Principles of Programming Languages}
\def\conficsme{Proceedings of the International Conference on Software Maintenance and Evolution}
\def\conffsesymposium{Proceedings of the International Symposium on Foundations of Software Engineering}
\def\conffsemeeting{Proceedings of the Joint Meeting on Foundations of Software Engineering}
\def\conficsm{Proceedings of the International Conference on Software Maintenance}
\def\confcgit{Proceedings of the International Conference on Computer Graphics and Interactive Techniques}
\def\confaaai{Proceedings of the AAAI Conference on Artificial Intelligence}
\def\confchi{Proceedings of the CHI Conference on Human Factors in Computing Systems}
\def\confemnlp{Proceedings of the Conference on Empirical Methods in Natural Language Processing}
\def\confiui{Proceedings of the International Conference on Intelligent User Interfaces}
\def\confuist{Proceedings of the Symposium on User Interface Software and Technology}
\def\confase{Proceedings of the International Conference on Automated Software Engineering}
\def\confissta{Proceedings of the International Symposium on Software Testing and Analysis}
\def\conffase{Proceedings of the International Conference on Fundamental Approaches to Software Engineering}
\def\conficpc{Proceedings of the International Conference on Program Comprehension}
\def\conficse{Proceedings of the International Conference on Software Engineering}
\def\confchase{Proceedings of the International Workshop on Cooperative and Human Aspects of Software Engineering}
\def\confacl{Proceedings of the Annual Meeting of the Association for Computational Linguistics}
\def\confvlhcc{Proceedings of the Symposium on Visual Languages and Human-Centric Computing}
\def\conflas{Proceedings of the Conference on Learning at Scale}
\def\confisese{Proceedings of the International Symposium on Empirical Software Engineering}
\def\confsigcse{Proceedings of the Technical Symposium on Computer Science Education}
\def\confwcre{Proceedings of the Working Conference on Reverse Engineering}
\def\confhatra{Proceedings the Workshop on Human Aspects of Types and Reasoning Assistants}
\def\confonward{Proceedings the Symposium on New Ideas in Programming and Reflections on Software}
\def\confvldb{Proceedings of the International Conference on Very Large Data Bases}
"}

@article{nicholson2021scite,
  title={scite: A smart citation index that displays the context of citations and classifies their intent using deep learning},
  author={Nicholson, Josh M and Mordaunt, Milo and Lopez, Patrice and Uppala, Ashish and Rosati, Domenic and Rodrigues, Neves P and Grabitz, Peter and Rife, Sean C},
  journal={Quantitative science studies},
  volume={2},
  number={3},
  pages={882--898},
  year={2021},
  publisher={MIT Press One Rogers Street, Cambridge, MA 02142-1209, USA journals-info~…}
}

@article{memon2024search,
  title={Search engines post-ChatGPT: How generative artificial intelligence could make search less reliable},
  author={Memon, Shahan Ali and West, Jevin D},
  journal={arXiv preprint arXiv:2402.11707},
  year={2024}
}

@inproceedings{venkit2024search,
  title={Search engines in the AI era: A qualitative understanding to the false promise of factual and verifiable source-cited responses in LLM-based search},
  author={Venkit, Pranav Narayanan and Laban, Philippe and Zhou, Yilun and Mao, Yixin and Wu, Chien-Sheng},
  booktitle={Proceedings of the 2025 ACM Conference on Fairness, Accountability, and Transparency},
  pages={1325--1340},
  publisher={ACM},
  year={2025}
}

@article{august2023paper,
  title={Paper Plain: Making medical research papers approachable to healthcare consumers with natural language processing},
  author={August, Tal and Wang, Lucy Lu and Bragg, Jonathan and Hearst, Marti A and Head, Andrew and Lo, Kyle},
  journal={ACM Transactions on Computer-Human Interaction},
  volume={30},
  number={5},
  publisher={ACM},
  articleno={74},
  year={2023},
  publisher={ACM New York, NY}
}

@inproceedings{fok2023scim,
  title={Scim: Intelligent skimming support for scientific papers},
  author={Fok, Raymond and Kambhamettu, Hita and Soldaini, Luca and Bragg, Jonathan and Lo, Kyle and Hearst, Marti and Head, Andrew and Weld, Daniel S},
  booktitle={Proceedings of the 28th International Conference on Intelligent User Interfaces},
  pages={476--490},
  year={2023},
  publisher={ACM}
}

@manual{perplexity,
  url = {https://perplexity.ai/},
  title = {Perplexity},
  note = {Last accessed 31 July 2026}
}

@manual{copilot,
  title = {Microsoft Copilot},
  url = {https://copilot.microsoft.com/},
  note = {Last accessed 31 July 2026}
}

@manual{you,
  url = {https://you.com},
  title = {you.com},
  note = {Last accessed 31 July 2026}
}

@manual{papercraft,
    url = {https://openreader.semanticscholar.org/PaperCraft},
    title = {PaperCraft},
    note = {Last accessed 31 July 2026}
}

@manual{pypdf,
    url = {https://pypi.org/project/pypdf/},
    title = {pypdf},
    note = {Last accessed 31 July 2026}
}

@manual{chatdoc,
    url = {https://chatdoc.com/},
    title = {chatdoc.com},
    note = {Last accessed 31 July 2026}    
}

@manual{deepresearch,
  url = {https://openai.com/index/introducing-deep-research/},
  title = {OpenAI Deep Research},
  note = {Last accessed 31 July 2026}
}

@article{lindemann2024chatbots,
  title={Chatbots, search engines, and the sealing of knowledges},
  author={Lindemann, Nora Freya},
  journal={AI \& SOCIETY},
  pages={5063--5076},
  year={2024},
  publisher={Springer}
}

@inproceedings{sharma2024generative,
  title={Generative echo chamber? Effect of LLM-powered search systems on diverse information seeking},
  author={Sharma, Nikhil and Liao, Q. Vera and Xiao, Ziang},
  booktitle={Proceedings of the 2024 CHI Conference on Human Factors in Computing Systems},
  publisher={ACM},
  articleno={1033},
  year={2024}
}

@inproceedings{head2021augmenting,
  title={Augmenting scientific papers with just-in-time, position-sensitive definitions of terms and symbols},
  author={Head, Andrew and Lo, Kyle and Kang, Dongyeop and Fok, Raymond and Skjonsberg, Sam and Weld, Daniel S and Hearst, Marti A},
  booktitle={Proceedings of the 2021 CHI Conference on Human Factors in Computing Systems},
  publisher={ACM},
  articleno={413},
  year={2021}
}

@inproceedings{fok2024marco,
  title={Marco: Supporting Business Document Workflows via Collection-Centric Information Foraging with Large Language Models},
  author={Fok, Raymond and Lipka, Nedim and Sun, Tong and Siu, Alexa F},
  booktitle={Proceedings of the 2024 CHI Conference on Human Factors in Computing Systems},
  articleno={842},
  publisher={ACM},
  year={2024}
}

@inproceedings{liu2024selenite,
  title={Selenite: Scaffolding Online Sensemaking with Comprehensive Overviews Elicited from Large Language Models},
  author={Liu, Michael Xieyang and Wu, Tongshuang and Chen, Tianying and Li, Franklin Mingzhe and Kittur, Aniket and Myers, Brad A},
  publisher={ACM},
  booktitle={Proceedings of the 2024 CHI Conference on Human Factors in Computing Systems},
  articleno={837},
  year={2024}
}

@article{Kinney2023TheSS,
  title={The Semantic Scholar Open Data Platform},
  author={Rodney Michael Kinney and Chloe Anastasiades and Russell Authur and Iz Beltagy and Jonathan Bragg and Alexandra Buraczynski and Isabel Cachola and Stefan Candra and Yoganand Chandrasekhar and Arman Cohan and Miles Crawford and Doug Downey and Jason Dunkelberger and Oren Etzioni and Rob Evans and Sergey Feldman and Joseph Gorney and David W. Graham and F.Q. Hu and Regan Huff and Daniel King and Sebastian Kohlmeier and Bailey Kuehl and Michael Langan and Daniel Lin and Haokun Liu and Kyle Lo and Jaron Lochner and Kelsey MacMillan and Tyler C. Murray and Christopher Newell and Smita R Rao and Shaurya Rohatgi and Paul Sayre and Zejiang Shen and Amanpreet Singh and Luca Soldaini and Shivashankar Subramanian and A. Tanaka and Alex D Wade and Linda M. Wagner and Lucy Lu Wang and Christopher Wilhelm and Caroline Wu and Jiangjiang Yang and Angele Zamarron and Madeleine van Zuylen and Daniel S. Weld},
  journal={ArXiv},
  year={2023},
  volume={abs/2301.10140},
}

@inproceedings{lo2023papermage,
  title={PaperMage: a unified toolkit for processing, representing, and manipulating visually-rich scientific documents},
  author={Lo, Kyle and Shen, Zejiang and Newman, Benjamin and Chang, Joseph Z and Authur, Russell and Bransom, Erin and Candra, Stefan and Chandrasekhar, Yoganand and Huff, Regan and Kuehl, Bailey and others},
  booktitle={Proceedings of the 2023 Conference on Empirical Methods in Natural Language Processing: System Demonstrations},
  pages={495--507},
  publisher={Association for Computational Linguistics},
  year={2023}
}

@inproceedings{duck2025finding,
  title={Finding Needles in Document Haystacks: Augmenting Serendipitous Claim Retrieval Workflows},
  author={D{\"u}ck, Moritz and Holter, Steffen and Chan, Robin Shing Moon and Sevastjanova, Rita and El-Assady, Mennatallah},
  booktitle={Proceedings of the 2025 CHI Conference on Human Factors in Computing Systems},
  articleno={1003},
  publisher={ACM},
  year={2025}
}

@inproceedings{pirolli1995information,
  title={Information foraging in information access environments},
  author={Pirolli, Peter and Card, Stuart},
  booktitle={Proceedings of the SIGCHI Conference on Human Factors in Computing Systems},
  publisher={ACM},
  pages={51--58},
  year={1995}
}

@inproceedings{Singh2025Ai2SQ,
    title = "Ai2 Scholar {QA}: Organized Literature Synthesis with Attribution",
    author = "Singh, Amanpreet  and
      Chang, Joseph Chee  and
      Haddad, Dany  and
      Naik, Aakanksha  and
      Hwang, Jena D.  and
      Kinney, Rodney  and
      Weld, Daniel S  and
      Downey, Doug  and
      Feldman, Sergey",
    editor = "Mishra, Pushkar  and
      Muresan, Smaranda  and
      Yu, Tao",
    booktitle = "Proceedings of the 63rd Annual Meeting of the Association for Computational Linguistics (Volume 3: System Demonstrations)",
    month = jul,
    year = "2025",
    publisher = "Association for Computational Linguistics",
    pages = "513--523",
    ISBN = "979-8-89176-253-4"
}

@inproceedings{cohan-etal-2019-structural,
    title = "Structural Scaffolds for Citation Intent Classification in Scientific Publications",
    author = "Cohan, Arman  and
      Ammar, Waleed  and
      van Zuylen, Madeleine  and
      Cady, Field",
    editor = "Burstein, Jill  and
      Doran, Christy  and
      Solorio, Thamar",
    booktitle = "Proceedings of the 2019 Conference of the North {A}merican Chapter of the Association for Computational Linguistics: Human Language Technologies",
    month = jun,
    year = "2019",
    publisher = "Association for Computational Linguistics",
    pages = "3586--3596"
}

@inproceedings{lauscher-etal-2018-argument,
    title = "An Argument-Annotated Corpus of Scientific Publications",
    author = "Lauscher, Anne  and
      Glava{\v{s}}, Goran  and
      Ponzetto, Simone Paolo",
    editor = "Slonim, Noam  and
      Aharonov, Ranit",
    booktitle = "Proceedings of the 5th Workshop on Argument Mining",
    month = nov,
    year = "2018",
    publisher = "Association for Computational Linguistics",
    pages = "40--46"
}

@article{asai2026synthesizing,
  author  = {Asai, Akari and He, Jacqueline and Shao, Rulin and Shi, Weijia and
             Singh, Amanpreet and Chang, Joseph Chee and Lo, Kyle and
             Soldaini, Luca and Feldman, Sergey and D'Arcy, Mike and
             Wadden, David and Latzke, Matt and Sparks, Jenna and
             Hwang, Jena D. and Kishore, Varsha and Tian, Minyang and
             Ji, Pan and Liu, Shengyan and Tong, Hao and Wu, Bohao and
             Xiong, Yanyu and Zettlemoyer, Luke and Neubig, Graham and
             Weld, Daniel S. and Downey, Doug and Yih, Wen-tau and
             Koh, Pang Wei and Hajishirzi, Hannaneh},
  title   = {Synthesizing scientific literature with retrieval-augmented language models},
  journal = {Nature},
  volume  = {650},
  number  = {8103},
  pages   = {857--863},
  year    = {2026},
  doi     = {10.1038/s41586-025-10072-4},
}

% \IfFileExists{main.bbl}{
%   \input{main.bbl} % arXiv uses the frozen reference list
% }{
%   \bibliographystyle{ACM-Reference-Format}
%   \bibliography{refs,andrew-base,andrew-extras} % Overleaf uses .bib if .bbl missing
% }

\appendix

\section{Appendix}

\begin{changes}
\subsection{Implementation details}
\label{sec:implementation-details}

Our implementation is a proof of concept combining off-the-shelf AI components with some human supervision. The pipeline is too slow for deployment in existing answer engines. This was an intentional tradeoff we made so that our study could better characterize the upper bound of the design's utility. We used the strongest general-purpose model available to us at the time of development, GPT-5.1, in all stages of the pipeline. Prompts appear in Appendix~\ref{sec:pipeline-prompts}.

\paragraph{Fetching answers}
We query OpenSciLM's web API~\cite{asai2026synthesizing}, which generates attributed answers over 45 million open-access papers and returns answers along with citations that have unique IDs for looking up cited papers.

\paragraph{Extracting claims}
For each sentence containing a citation, we prompt the LLM to return minimal, testable scientific claims relating to the sentence's main point.

\paragraph{Extracting excerpts}
We fetch PDFs for cited papers via open-access links from the Semantic Scholar API~\cite{Kinney2023TheSS} and extract their text with pyPDF~\cite{pypdf}. We prompt the LLM with this text to extract evidence excerpts relating to each claim, classify each excerpt according to degree and valence, and generate an explanation to help make the passage understandable out of context.

\paragraph{Fetching second-degree sources}
We resolve references to sources within second-degree excerpts using Asta text snippet search, or, failing that, by extracting relevant reference strings with an LLM  and matching them to documents with the Semantic Scholar API. Some sources could still not be resolved automatically; we looked up these references manually and downloaded their PDFs when not paywalled. This covered 74.5\% of sources referenced in second-degree excerpts. Given the large number of second-degree sources to process, we opted for a more rapid excerpt extraction prompt for these sources, which limited the number of extracted excerpts to three per source.

\paragraph{UI}
We processed PDFs to add interactions and annotations offline before study sessions. To highlight excerpts in source PDFs, we located them with PaperMage~\cite{lo2023papermage} by matching high-overlap token sequences. Issues arose when PDFs consisted of images of text rather than text itself, leading to the exclusion of one PDF, or when the LLM-extracted evidence was not verbatim. We found that passing the full text into GPT-5.1 and prompting it to extract verbatim excerpts mitigated this latter issue. We added UI components from the PaperCraft~\cite{papercraft} library to the PDF to support scrolling to and highlighting the detected passages.
\end{changes}

\begin{changes}
\subsection{Main LLM prompts used in attribution gradients pipeline}
\label{sec:pipeline-prompts}

All prompts below were submitted to GPT-5.1 with structured output format (Pydantic/JSON-schema). Source papers were chunked into 8{,}000--10{,}000 character segments to fit context limits.

\subsubsection{Evidence extraction}
\label{sec:prompt-evidence-full}

Used to identify all passages in a source paper that support or contradict a given claim.

\begin{lstlisting}
**TASK: COMPREHENSIVE EVIDENCE EXTRACTION**

Conduct a thorough analysis of the research paper below to extract ALL passages that provide credible evidence for or against the given statement. Apply rigorous scientific standards throughout your analysis.

**EXTRACTION REQUIREMENTS:**

1. **Completeness**: Review the entire text systematically. Do not skip any sections including abstract, methods, results, discussion, or conclusions.

2. **Verbatim Accuracy**: Extract passages exactly as written in the source, including:
   - All punctuation, parentheticals, and formatting
   - Complete citation strings (e.g., "[2]", "(Smith et al., 2023)")
   - Statistical notations, p-values, effect sizes, confidence intervals
   - Technical terminology and abbreviations as written

3. **Evidence Standards**: Only extract content that constitutes legitimate scientific evidence:
   - **INCLUDE**: Empirical results, statistical findings, experimental outcomes, systematic observations, meta-analysis results, validated measurements
   - **INCLUDE**: Citations to established research findings when relevant to the statement
   - **EXCLUDE**: Hypotheses, research objectives, methodology descriptions (unless methodology itself is the subject of evaluation), author opinions, speculation

4. **Passage Length**: Typically one sentence, but may extend to multiple sentences if required for:
   - Complete statistical reporting (e.g., including confidence intervals, effect sizes)
   - Coherent presentation of complex findings
   - Preservation of essential context

**EVIDENCE CLASSIFICATION:**

For each extracted passage, provide these annotations:

* **Valence**:
  - "support": Evidence clearly supports or aligns with the statement
  - "contradict": Evidence clearly opposes or contradicts the statement
  - Apply strict standards, ambiguous or weak evidence should be excluded entirely

* **Evidence Kind**:
  - "primary": Original findings, data, or analyses presented for the first time in this paper
  - "secondary": References to established findings from prior published research

* **Explanation**: Provide essential context in clear, accessible language:
  - Define technical terms or statistical concepts
  - Explain the relevance to the statement
  - Clarify what the evidence demonstrates
  - Use simple, direct sentences appropriate for a general academic audience

Output the extracted passages and annotations as a JSON object with this format. This example assumes an input statement of "Eating eggs accelerates muscle growth.":

{
  "passages": [
    {"passage": "Prior studies have shown that consumption of eggs led to an increase in volume of the deltoids [1][2].", "evidence_kind": "secondary", "valence": "support", "explanation": "Deltoids are a muscle group. They are sometimes measured in studies of muscle growth."},
    {"passage": "Our ANOVA showed no significant effect (p = .31).", "evidence_kind": "primary", "valence": "contradict", "explanation": "The test looked for different muscle growth for egg eaters. The p-value is high. This suggests the difference was a fluke. So there is no evidence here for eggs affecting muscle growth."}
  ]
}

Statement: {STATEMENT}
Text: {TEXT}
\end{lstlisting}

\subsubsection{Claim Decomposition}
\label{sec:prompt-subclaims}

Used to break AI-generated answer summaries into atomic, independently verifiable claims.

\begin{lstlisting}
**TASK: SYSTEMATIC CLAIM EXTRACTION AND DECOMPOSITION**

Analyze the provided AI-generated research summary to identify and decompose verifiable scientific claims that require evidence validation.

**STEP 1: SENTENCE IDENTIFICATION**
Parse the summary into individual sentences. Focus exclusively on sentences that:
- Contain explicit citation markers (e.g., "[1]", "[3]", "(Author et al., Year)")
- Present factual assertions that can be empirically verified
- Include substantive information relevant to the research question
- Make claims that would benefit from validation through cited source materials

**STEP 2: CLAIM DECOMPOSITION CRITERIA**
For each qualifying sentence, extract minimal, testable scientific claims according to these principles:

1. **Atomicity**: Each claim should test exactly one scientific proposition
2. **Testability**: Claims must be empirically verifiable through research evidence
3. **Clarity**: Use precise, unambiguous language avoiding compound or complex structures
4. **Relevance**: Claims should directly relate to the core assertions of the sentence
5. **Uniqueness**: Avoid duplicate or highly similar claims

**CONSTRAINTS:**
- Maximum 6 claims per sentence (typically 1-3)
- Claims must be complete, standalone statements
- Preserve scientific accuracy from the original text
- Maintain appropriate level of specificity

**QUALITY STANDARDS:**
- **INCLUDE**: Specific research findings, quantitative relationships, causal claims, comparative statements, methodological assertions
- **EXCLUDE**: Vague generalizations, obvious statements, purely definitional content, speculative language

**EXAMPLE TRANSFORMATION:**
Original: "Daily consumption of omega-3 fatty acids significantly improves cognitive performance and reduces inflammatory markers in elderly adults [2][5]."

Decomposed Claims:
- "Daily consumption of omega-3 fatty acids significantly improves cognitive performance in elderly adults."
- "Daily consumption of omega-3 fatty acids reduces inflammatory markers in elderly adults."

**RESEARCH CONTEXT:**
Question: {QUESTION}

**TEXT TO ANALYZE:**
{TEXT}
\end{lstlisting}
\end{changes}

\subsection{Accuracy of evidence extraction for stimuli}
\label{sec:snapshot}

While the implementation is not a contribution of this work, we briefly describe the accuracy of the evidence extraction and classification shown to users in the study. This helps contextualize what users were seeing in the study---evidence that was largely extracted and classified correctly, but not perfectly.
% We audited the outputs of our implementation to understand the extent to which attribution gradients showed appropriate excerpts, appropriately organized.
We recruited an HCI Ph.D. student with experience reading in the tasks' domains to examine all 113 evidence excerpts participants might see across the two tasks and tag their relationship to the claim.\footnote{This omitted unraveled excerpts from second-degree sources, but did include second-degree excerpts (those containing citations to other papers).} Of excerpts that the judge tagged as supporting, 73.5\% were correctly classified by our pipeline as such; of those classified otherwise, the vast majority (92.6\%) were considered by the judge as broadly relevant to the claim but lacking specific evidence. All excerpts classified as contradicting were judged to be correctly classified. Classification into evidence kind was more mixed: for instance, what was classified as first-degree supporting evidence was judged to be evidence consisting of original data or results 36.4\% of the time; theoretical arguments, synthesis, or interpretation of facts 40.0\% of the time (which can sometimes be considered first-degree evidence), otherwise broadly relevant 20.0\% of the time, and unjustified assertions 3.6\% of the time. For second-degree supporting evidence, the judge reported that the evidence referenced other sources 46.8\% of the time, unjustified assertions 12.8\% of the time, and broadly relevant information 29.8\% of the time. In other words, valence was largely correct and degree had more mixed results.

\subsection{Effort and questions in the baseline}
\label{sec:baseline}

% \begin{figure}[H]
%     \centering
%     \includegraphics[width=\linewidth]{source/figures/chatdoc.pdf}
%     \caption{The baseline interface. \mdseries ChatDoc is a chat-with-papers interface that allows users to ask questions of uploaded documents and receive answers with links into the source text.\andrew{I hate to say this but this image may be copyrighted material. Maybe we can get a quick snapshot of the interface without it.}}
%     \label{fig:chatdoc}
% \end{figure}

% \subsection{Effort and questions in the baseline}

The baseline condition involved some kinds of work not present in the attribution gradients condition. Participants often engaged with documents through the conversational interface of ChatDOC, asking on average 2.4 questions. Some of the most common kinds of questions were:
\begin{itemize}
  \item Information extraction and summarization ($n=5$), such as ``Provide the top 10 takeaways from these studies and what they say about how recommendation algorithms affect the formation of users' preferences''
  \item Clarification questions ($n=5$), such as ``Did the paper provide a control of the user's preference before and after their intervention?''
  \item Requests for contradiction ($n=2$), such as ``How would someone might point out the flaws in these experiments setup to make counterargument about the claim?''
\end{itemize}

\begin{changes}
Participants also opened the source PDFs, though less often than with attribution gradients, opening a median of 2 papers (M = 1.7). Participants' accounts suggested the chat interface was valued for orientation and also as a substitute for reading. P8 found the baseline ``really good for the AI chatbot for an overview without going through details.'' P5 noted ``I found myself using the AI chatbot a little too much,'' describing how scrolling through generated answers ``funneled me into agreement.'' P12 reflected that with the baseline ``I never really understood what the main point was.''

Moving between answer, sources, and chat could take time. One participant spent 47 seconds on a single pass: reading through the generated answer, clicking into part of it, opening the source PDF, and then asking ChatDOC a question to make sense of what they had read. Reading the results section of one paper took the same participant 6.5 minutes of the 10-minute task. Some participants also asked ChatDOC for the kinds of information attribution gradients provided: 2 asked for contradictory evidence and 1 asked where the chatbot's answers came from.

\end{changes}
\vspace{-4ex}
\end{document}